# Towards *In Silico* Mining for Superconductors – Cutting the Gordian Knot

Vedad Babic[a,b] and Itai Panas[a]

[a]Department of Chemistry and Chemical Engineering, Chalmers University of Technology, 412 96 Gothenburg, Sweden

[b]Corresponding author: vedadb@chalmers.se

**Abstract**

A random forest regression based supervised machine learning method to predict experimental critical temperature of superconductivity from the electronic band structure, as obtained from Density Functional Theory, is demonstrated. This complementarity between experiment and theory draws inspiration from the merging of Kohn-Sham and Bogoliubov-De Gennes equations [W. Kohn, W, EKU Gross, and LN Oliveira, *Int. J. of Quant. Chem.*, **36**(23), 611-615 (1989)]. Features in the Kohn-Sham Density Functional Theory band structure away from $E_F$ becoming decisive for the superconducting gap demonstrates this divide-and-conquer physical understanding. Not committing to any microscopic mechanism for the SC at this stage, it implies that in different classes of materials, different electronic features are responsible for the superconductivity. However, training on known members of a class, the performance of new members may be predicted. The method is validated for the A15 materials, including both binary $A_3X$ and ternary $A_6XY$ intermetallics, A=V, Nb, demonstrating that the two do indeed belong to the same class of superconductors.

**Introduction**

Superconductivity (SC) is a state of matter that is as robust as it is elusive. Its potential technological importance is matched only by our fascination for the phenomenon. Correspondingly, titanic efforts have been undertaken over the years in formulating and exploring possible principles guiding the search for new superconducting materials. Groundbreaking complementary contextual, conceptual and phenomenological approaches [1-3] were developed, focusing on different aspects of the phenomenon. Challenging in this endeavor is that the phenomenon is confined to a narrow energy window in vicinity of the Fermi energy, where mixing of hole- and particle states takes place, as reflected in the

Cooper pairing. Experimental studies on fundamental characteristics have served to establish and consolidate this understanding further [4,5] leading the way to technological applications.

Thus, superconductivity results from entanglements among states across $E_F$ that serve to suppress competing resistive thermal excitations. For a material to host the SC, the fine interplay between hole- and electron states must be protected by resilient atomic structural elements or else spontaneous crystal symmetry breaking will cause lifting of the required electronic degeneracy at $E_F$. In fact, residual protected degeneracies at $E_F$ are understood to result owing to structural relaxations where lifting some degeneracies leaves others protected. As a rule, ambiguity remains whether the said resulting features will correlate constructively to produce the SC, or if any competing electronic order will prevail, or if SC will emerge from coopetition between competing orders.

As of today, there is a mismatch between our quantum mechanical description of materials that harbor the superconductivity, and the details that unambiguously lead to the phenomenon. Consequently, as of today, any claimed prediction of new superconducting materials based on $1^{st}$ principles alone must be deemed accidental, leaving us with traditional empirics-based serendipity or proposing new compounds based on essential similarities with already known superconductors. Such classes of superconductors include the cuprates [6], the layered iron-based materials [7], and the A15 intermetallics [8]. Currently, a new class of hydride based materials exhibiting possible room-temperature superconductivity albeit at very high pressures ~100-200 GPa are being explored experimentally [9,10] as well as by means of physics based machine learning [11] guided by the Eliashberg spectral function for electron-phonon coupling. Yet, it was shown in a recent machine learning study – based on 145 experimental element specific materials features [12] – that predictability for one class of superconductors was not transferable to other classes. Hence, while the SC phenomenon is generic, albeit of either type I and type II, the essential materials features and the way they cooperate is unique to each class of superconducting materials.

Even if constrained to apply within individual structural classes, still, supervised machine learning could provide a powerful predictive tool in identifying and ranking yet missing members in the same class. Here we explore the extent to which $1^{st}$ principles electronic band structures – as obtained from Kohn-Sham Density Functional Theory (KS DFT) calculations – can replace all the said 145 experimental element specific materials features in the supervised machine learning [12]. In as much as this approach remains unbiased concerning the mechanism for the superconductivity, it leaves room for further developments in forthcoming studies. Here, conceptual, yet intuitive, justification for such a 1st principles-augmented supervised machine learning method is highlighted and demonstrated. Indeed, considering that top and bottom branches of bands that cross $E_F$ are commonly unproblematic for KS DFT, by interpolation between the said branches, the crystal orbitals in vicinity of $E_F$ in turn become a consequence of the continuity of the band. The remaining crucial ambiguity is owing to the *a priori* single-particle Kohn Sham crystal orbitals representation. Were the electron-electron and electron-phonon inter- and intra-band virtual scatterings to be considered explicitly, the said orbitals would acquire non-integer occupations. Fundamental work in line with this understanding has been undertaken previously, including solving the Bogoliubov-de Gennes (BdG) equations [13] in the framework of KS DFT [14,15]. It was emphasized by Kohn et al [16] that for an all-electronic mechanism for superconductivity, e.g., owing to Friedel oscillations [17], BdG-KS equations analogous to KS DFT would result, while in case of electron-phonon coupling the functional becomes system dependent, i.e., non-universal. A multi-band all-electronic realization, see [18-20], emerges from an multiconfigurational representation of the electron-electron interaction [21,22], whereby local near-degeneracies are included explicitly in the truncated variational Configuration Interaction wave function. To date, while consolidating our basic understanding, such efforts have had no predictive impact in search of new superconducting materials.

From [12] and [16] it is inferred that *if* all members of a class of superconductors share a common mechanism of superconductivity, *then* emerging properties of their KS crystal orbitals may indeed serve descriptors of the SC, i.e., correlating with $T_C$. This would offer an alternative means to account for the crucial fine hole-particle interplay – missing in KS DFT [23] but inferred by the Hohenberg-Kohn theorem [24] – by replacing all-formal approaches by supervised learning, thereby associating KS DFT band structure features to experimental $T_C$ measurements. Repeatedly, for non-pathologic cases, it would follow that unknown members of a class of superconductors could be predicted from the known members of this class from their KS DFT band structures.

To test the validity of this inference, *firstly*, a sufficiently large class of superconducting materials is required for efficient training, *secondly*, the applicability of KS DFT within the generalized gradient approximation GGA should be uncontested for the material class considered, and *thirdly,* the ability to predict materials that extend the class should be conclusive.

These requirements are fulfilled for the archaic vanadium and niobium based quasi-one-dimensional A15 $A_3B$ class of superconductors, A=V or Nb [25]. These systems have been studied extensively by means of electronic structure calculations over the years [26-28], and some essential properties are shown in Figure 1. Thus, the A-sublattice is understood to form interwoven chains of V or Nb. A near-degenerate manifold of states at $E_F$ results, originating mainly from the fivefold atomic d-orbitals, where the two atomic $d_\delta$-orbitals oriented orthogonal to the individual chains' directions become separated from the $d_\sigma$ and the two $d_\pi$ components. Characteristic is the subdivision of the partial density of states owing to this d-manifold into $d^+$ and $d^-$ bundles, where $E_F$ resides at the top of the $d^+$ bundle. The decisive interchain coupling to switch on and attenuate the $T_C$ for the SC, however, is controlled by the B element. One effect is owing to the B elements in the group 13 and 14, placing a set of coalescing bands close to $E_F$, in vicinity of the R point. This effect originates from the p-character of these elements [29], cf. Figure 1 again.

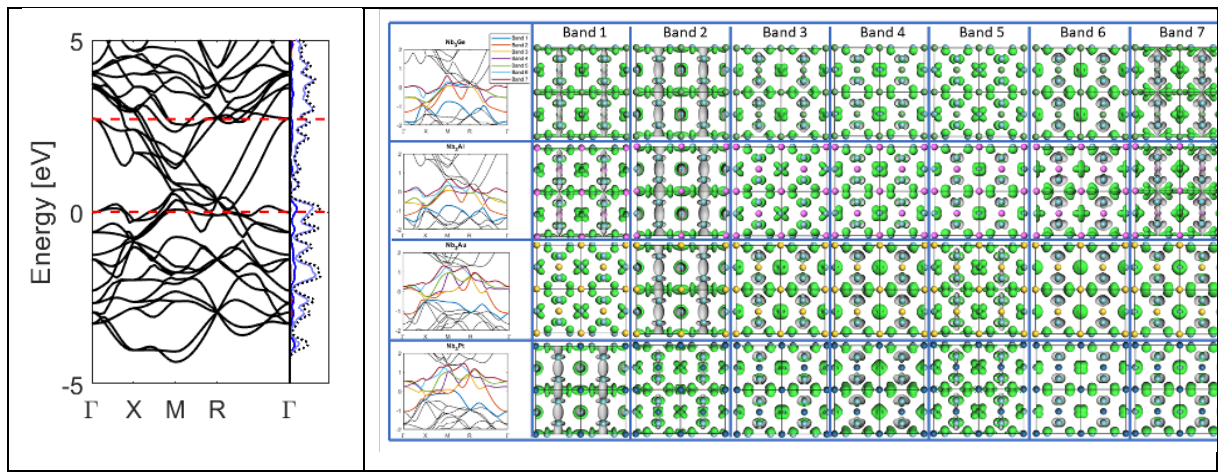

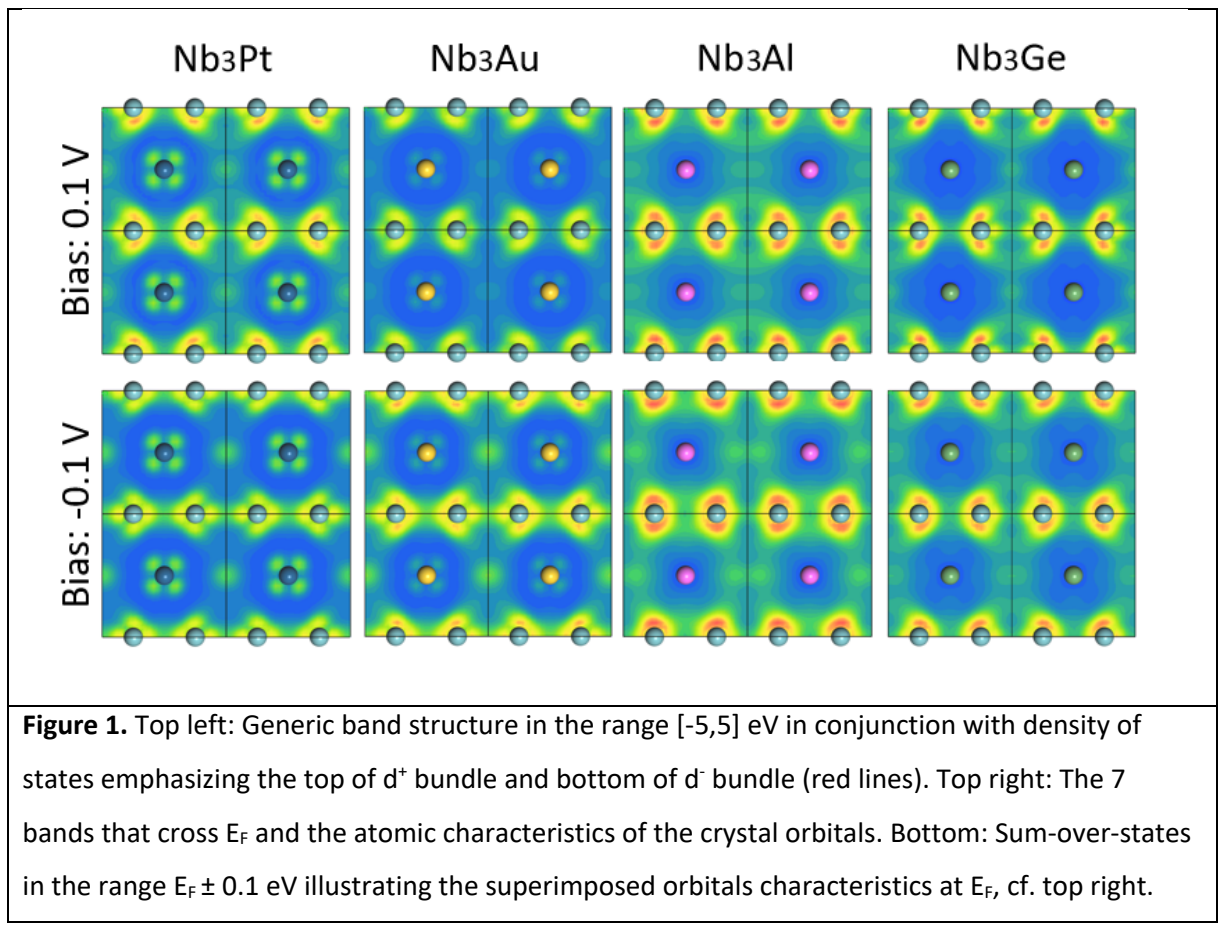

**Figure 1.** Top left: Generic band structure in the range [-5,5] eV in conjunction with density of states emphasizing the top of $d^+$ bundle and bottom of $d^-$ bundle (red lines). Top right: The 7 bands that cross $E_F$ and the atomic characteristics of the crystal orbitals. Bottom: Sum-over-states in the range $E_F \pm 0.1$ eV illustrating the superimposed orbitals characteristics at $E_F$, cf. top right.

In what follows, the ability to associate experimental $T_C$s to computed band structures by means of supervised machine learning is demonstrated. Accuracy of models trained on three different subsets of the data is compared, the {$V_3X$} subset, the {$Nb_3X$} subset, and the joint {$Nb_3X; V_3X$} subset, respectively. The accuracy of a model trained on one subset to predict the $T_C$s of members of its own subset is evaluated by means of leave-one-out cross validation LOOCV. More crucially, the ability of a model, trained on one subset, to predict the $T_C$s of members of the remaining ternary and/or binary subsets, is reported. Pearson correlation coefficient $\rho(T_C^{pred}, T_C^{exp})$, Root Mean Square Error (RMSE) as well as Mean Absolute Error (MAE) are employed to quantify the performance. All results are summarized as heatmaps of the resulting $\rho(T_C^{pred}, T_C^{exp})$, MAE, and RMSE in Figure S1.

**Computational Method**

Binary and ternary Nb- and V-based A15 ($A_3B$) entries with reported $T_C$:s were extracted from the SuperCON database [30]. The maximum reported $T_C$ was taken as the nominal one for each compound. Subsets of the binary entries were reserved for training machine learning. Out of the total 26 binary entries included in the dataset, 13 are $Nb_3X$ and 13 $V_3X$. Moreover, the complete sets of V$_6$XY and Nb$_6$XY, 5 and 8 entries respectively, in the SuperCON database were included to evaluate the degree to which the proposed *1$^{st}$ principles* based supervised machine learning approach is valid.

It is inferred that A15 structures with higher long-range order correlate with higher T$_C$ [31]. Thus, the A$_6$B$_2$ unit cell was deemed appropriate in the electronic structure. Each composition was subject to spin-polarized DFT calculations employing the GGA PBE functional [32] and allowing for full geometry optimization. On the fly norm conserving pseudopotentials compatible with 1200 eV cut-off energy were employed to describe impact of core electrons, while the Brillouin zone was sampled with 0.05 Å$^{-1}$ k-points separation [33]. For this, the CASTEP code [34] within the Materials studio 6.0 suite [35] was employed. The convergence criteria included $2\cdot10^{-3}$ Å (displacement) and $2\cdot10^{-5}$ eV/atom (energy). The band structure was calculated using 0.01 Å$^{-1}$ k-point sampling separation along the $\Gamma \rightarrow X \rightarrow M \rightarrow R \rightarrow \Gamma$ trajectory.

Employing spline interpolation, all bands were sampled uniformly in k-space. Thus, the 1$^{st}$ Brillouin zone was sampled using 100 bins across the specified trajectory. For each k-bin the band structure was sampled in the ranges [-5,5] eV with 0.05 eV E-bin distance, i.e. 200 E-bins. The resulting k-resolved DOS (kDOS) was calculated by creating a 2D-histogram for the 100 x 200 bins (20 000 features), thereby rendering the repeated sampling of the band structures possible and manageable. It is noted that partitioning the band structure into bins renders the explicit association of a k-state with its band lost. Moreover, k-states associated with a bin in one system need not belong to the same bands as the k-states of other systems that are ascribed to the same bin. Any such implication should be "rediscovered" by the random forest regression.

To allow for the similarity of different band structures to emerge, smoothening by means of average (or mean) filtering was considered on the form

$$f_H^L(E_n, k_{n\prime}) = \frac{1}{H \cdot L} \sum_{i=n-\frac{H-1}{2}}^{n+\frac{H-1}{2}} \sum_{j=n\prime-\frac{L-1}{2}}^{n\prime+\frac{L-1}{2}} f_1^1(E_i, k_j) \quad (1)$$

Thereby, the new value of a feature becomes an average that includes its surrounding. Here, filtering on a square kernel of size H, i.e. H=L, were compared for H=1,3,5,...,13, cf. Figure 2. Other filtering strategies are left for future studies. Here, it is noted that the L-averaging renders the kDOS and DOS data increasingly similar and that the case H=L=1 corresponds to the original kDOS without filtering.

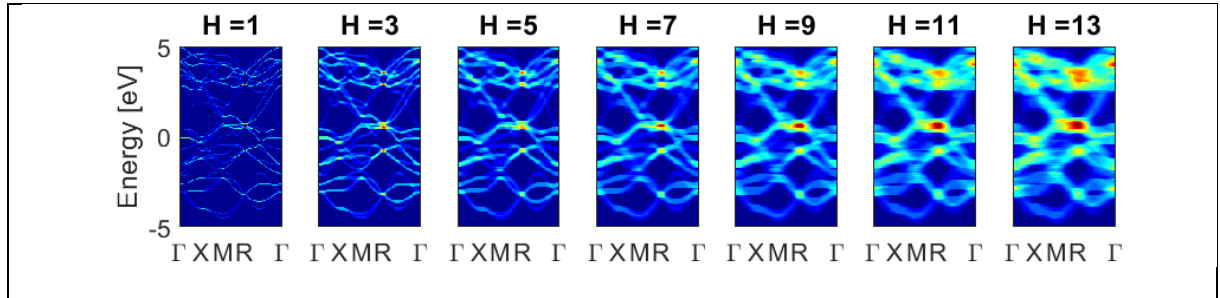

Figure 2. The effect of mean filtering on the kDOS of Nb$_3$Al using a kernel of size HxH, cf. Eq.1. Note that the case H=1 corresponds to unfiltered kDOS.

Dependence of model and prediction on reduction of features used in the kDOS by excluding those with an absolute Pearson correlation coefficient value, $|\rho(kDOS, T_C)|$ less than a threshold was investigated. The Pearson correlation coefficient is defined as:

$$\rho(X,Y) = \frac{cov(X,Y)}{\sigma(X)\sigma(Y)} = \frac{\sum_{i=1}^n (x_i - \bar{x})(y_i - \bar{y})}{\sqrt{\sum_{i=1}^n (x_i - \bar{x})^2} \sqrt{\sum_{i=1}^n (y_i - \bar{y})^2}}$$

Where n is the size of the subset on which the correlation is calculated. Note also that the correlation coefficient calculation is in fact performed separately for each bin in the kDOS i.e. $\rho(kDOS, T_C) = \rho(kDOS(E_i, k_j), T_C)$, c.f. Figure S8. The value of $\rho$ ranges between -1 and +1, where the former corresponds to perfect anti-correlation and the latter to perfect (positive) correlation, 0 corresponds to no correlation. Here, when reducing the feature set based on correlation it is performed on the absolute value of $\rho(kDOS, T_C)$ as both positive and negative correlations are in fact correlations. Another possibility would be to base the reduction on the square of $\rho$.

The resulting reduced feature set, which is calculated on the {Nb$_3$X,V$_3$X} subset, is visualized in Figure 3 while the full Pearson correlation maps, $|\rho(kDOS, T_C)|$, also for {Nb$_3$X} and {V$_3$X} are visualized in Figure S8.

The machine learning was performed using random forest regressor models with 10 trees from the Sci-Kit learn framework for Python.

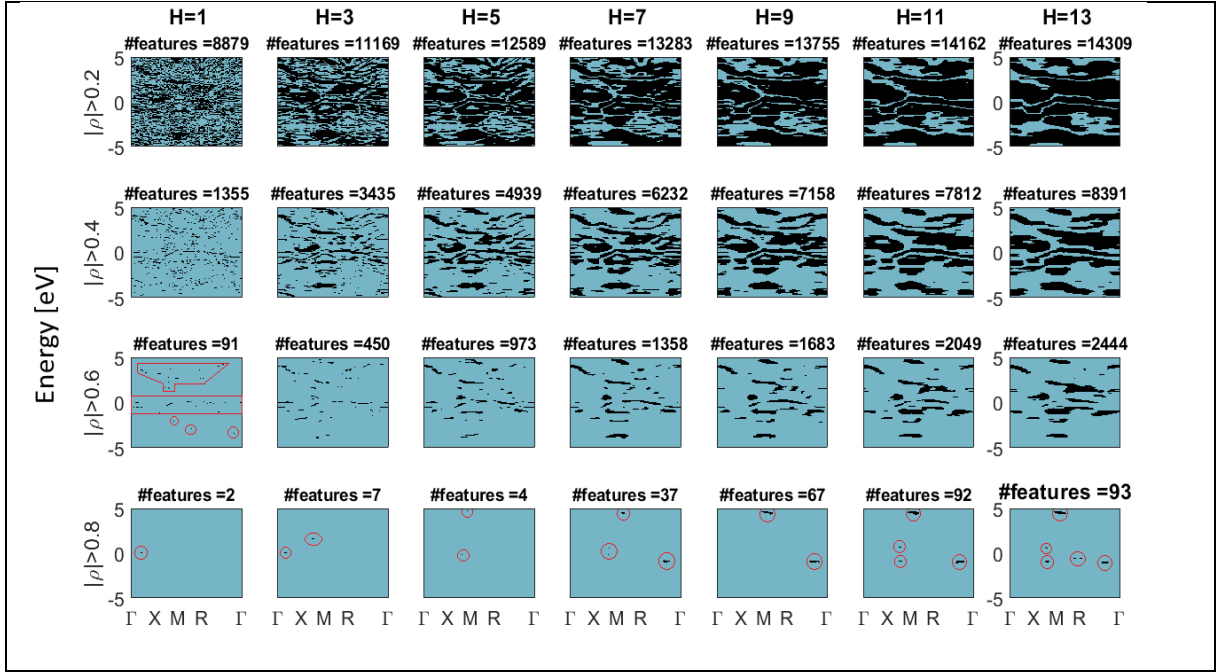

**Figure 3.** Pearson correlation maps based on all binaries, exhibiting the impact of reducing the feature set to those features that display correlation $|\rho(kDOS, T_C)| > 0.2$ to 0.8. Columns, left to right, show the impact of mean filtering using a square kernel of size H. Areas in black represent the remaining feature set (pixels) which fulfills the said criteria thereby forming basis for the threshold dependent correlation, and regression diagrams, see Figure 2 and Figures S2-S7. Where there are fewer than 100 features left, red encapsulation is used to highlight the features.

**Results and Discussion**

*Leave-one-out cross-validation:* The performances and internal consistencies of the models as well as the impact of filtering can be assessed by inspecting the results of LOOCV regressions for the three training subsets as shown in Figures S2-S4 and summarized in Figure 4a-c. It is noted that both mean filtering and feature reduction based on the absolute of the Pearson correlation coefficient $|\rho(kDOS, T_C)|$ independently improve on model performance. Also, while reducing the feature set to only relevant features do enhance the model, too few features may hamper the model training and performance, cf. Figure 4a-f, H=1, $|\rho(kDOS, T_C)| > 0.8$. Obviously, for $|\rho(kDOS, T_C)| > 0.8$, increasing the mean filtering tolerance from H=1 to H=13 renders the *a priori sparse* feature set increased as it allows for additional inferred similarity between band structures to emerge. This way, compounds that are *a priori* deemed dissimilar may acquire similarity, allowing for model performance to be tuned.

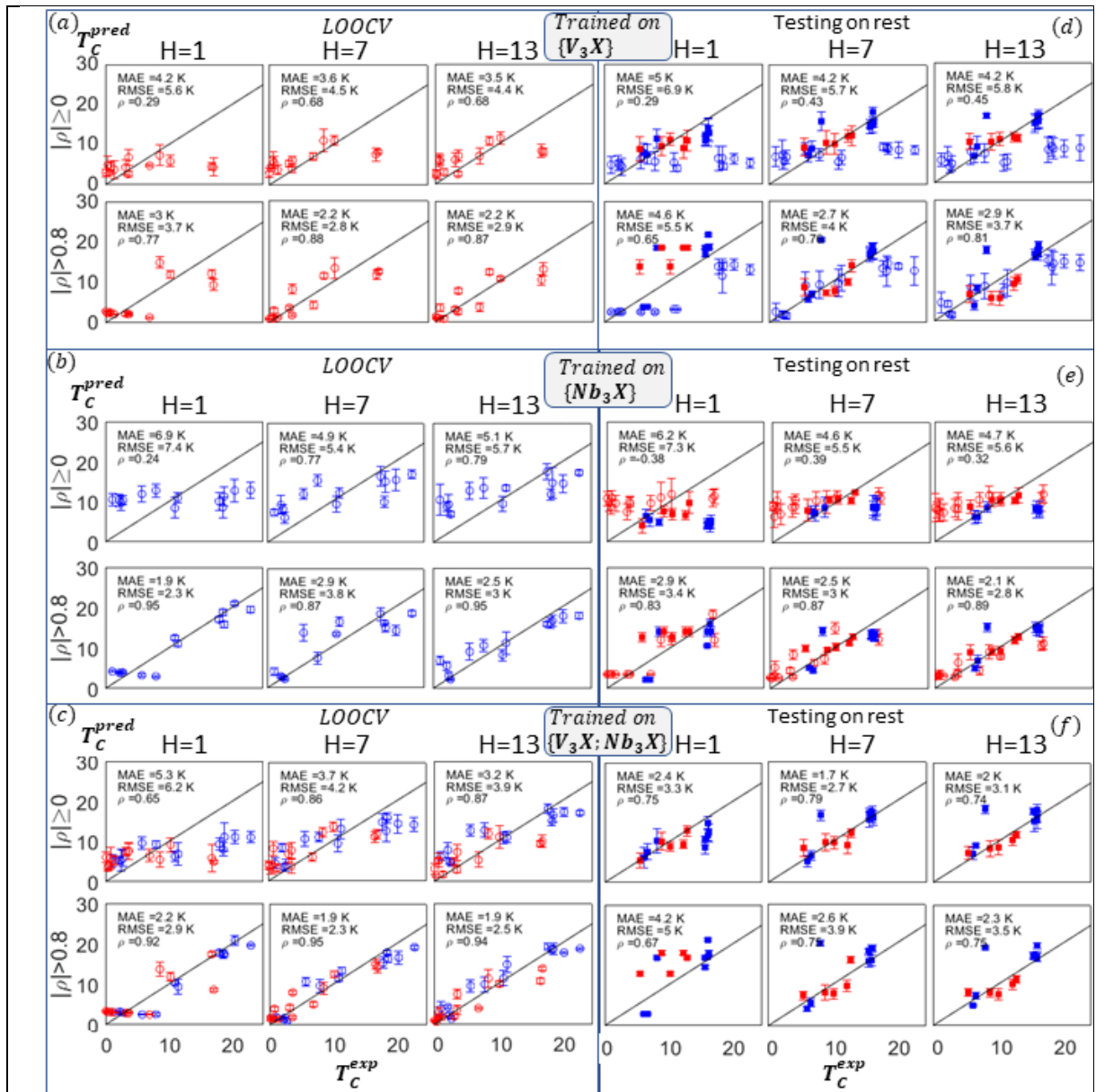

**Figure 4** Summary from the random forest regression analysis based on mean filtering and feature space reduction based on calculating the Pearson correlation coefficient, cf. Figures S2-S7. (a-c) Leave-one-out cross-validation LOOCV analysis, showing the ability to predict members within the subset based on training on all but one of the members. (d-f) Results of training on one subset of the data and testing on members of the remaining subsets. Rows: Impact on prediction owing to reduction of features to those that display *a priori* correlation, $|\rho(kDOS, T_C)| \geq 0$, that is no reduction: and $|\rho(kDOS, T_C)| > 0.8$. Columns: Impact of mean filtering on H×H adjacent elementary bins, H=1,7,13. Nb is blue, V is red.

***Testing on the rest:*** The ability of the 1$^{st}$ principles based random forest supervised machine learning method, when trained on all members of each of the three different subsets {V$_3$X}, {Nb$_3$X}, {Nb$_3$X;V$_3$X} to predict T$_C$s of members outside of the training subsets is shown in Figures S5-S7 and summarized in Figure 4d-f. For both {V$_3$X} and {Nb$_3$X} subsets, we find that irrespective of mean filtering, training on the unreduced kDOS feature set – $|\rho(kDOS, T_C)| \geq 0$ – upon predicting the rest, yields no impressive results. This holds for all H≤13, and this despite LOOCV on each subset showing fair results, cf Figure 4a-c. Rather, enhanced model accuracy is obtained

   a. by training on either binary subsystem after reducing the feature set to those features that show *a priori* correlation based on the full binary dataset c.f. Figure S8 for the difference in $|\rho(kDOS, T_C)|$ when calculated on the different subsets. Once feature set is screened on relevance, significant accuracy increase is obtained upon employing the mean filtering, cf. H=1 and H=13 for $|\rho(kDOS, T_C)| > 0.8$ in Figure 4d-e and Figures S6-S7.
   b. for all H on the unreduced $|\rho(kDOS, T_C)| \geq 0$ feature set, if both binary subsets are included in the training. This is owing to the complementarity of the two subsets, this enhancing co-correlated features while damping out others, see Figure 4f and Figure S5.

Indeed, it is repeatedly noted that reducing the feature set to include increasingly more relevant features, enhances the model performance only to a degree beyond which the model training becomes hampered by feature sparsity, cf. $|\rho(kDOS, T_C)| \geq 0$ and $|\rho(kDOS, T_C)| > 0.8$ for H=1 in Figure 4a-f and Figures S2-S7. Given this, it is truly remarkable how model performance is recovered and even surpassed upon applying the mean filtering, cf. $|\rho(kDOS, T_C)| > 0.8$ for H=1-13 in Figures S2-S7 and Figure 4 again.

It becomes particularly gratifying to note that – with one exception – the ability of the model that is trained on the all-binaries subset to predict the T$_C$s of the ternary A$_6$XY systems (see Figure 4f) is as good as it gets. In the case of Nb$_6$SiSn, experiment reports 8.3 K [36] while machine learning predicts ~20 K. This compound however is notoriously difficult to make [37], as is already the superconducting binary A15 Nb$_3$Si, which is metastable at ambient conditions [38,39]. While renewed attempts at measuring the T$_C$ of Nb$_6$SiSn is indeed in place, no sensational enhancement is expected beyond 20 K from the model.

Evidence to the fact that the two intermetallic systems {$V_3X$} and {$Nb_3X$} do indeed belong to the same class of superconductors is provided by this experiment augmented 1$^{st}$ principles method. This follows by inspecting the degrees to which the model that trains on $T_C$s and band structures of the {$Nb_3X$} subset is able to predict the $T_C$s of the {$V_3X$} subset and *vice versa*. Clearly, the band structure characteristics of the two subclasses correlate with SC in the same way. The slight asymmetry in efficiency of the two models, in predicting $T_C$s of the other, is due to the {$Nb_3X$} subset containing more compounds with higher $T_C$s than the {$V_3X$} subset, cf. Figures 4d & 4e and Figure S6-S7.

For a phenomenon that *per se* is inaccessible to KS DFT, the degree to which 1$^{st}$ principles kDOS based features provide decisive descriptors for how $T_C$ in the $A_3X$ intermetalics depends on choice of X element is truly remarkable. The striking effectiveness of the random forest regression based supervised machine learning method is taken to validate other models that employ KS DFT to describe essential aspects of the normal state of SC. Mutual predictive powers of the {$Nb_3X$} and {$V_3X$} training subsets, one predicting the other, is indeed promising. The higher quality of training on {$Nb_3X$} subset emphasizes yet again the importance of well-balanced data to support the training.

A main result of machine learning is the ability to put in question the validity of individual data. Here, we found that the 8.3 K $T_C$ reported for a material with nominal $Nb_6SnSi$ composition is not owing to superconductivity in the single-phase A15 structure of the same stoichiometry.

Finally, it is noted that the distribution of electronic band structure features correlating with $T_C$, and as determined from the corresponding Pearson correlation coefficients, do not single out the $E_F$ region. Crucially, this is taken to support the notion that once a class of superconductors has been identified, band structure signatures away from $E_F$, and thus possible to assess by KS DFT, become decisive for the tuning of properties at $E_F$, the superconducting gap included. This is consistent with the Hohenberg-Kohn theorem.

**In conclusion**, by providing proof-of-concept, this work hopes to pave the way for 1$^{st}$ principles *in silico* mining for superconductors, predicting new ones as well as reassessing failed ones. It opens up for a novel fruitful exploratory interplay between high-throughput electronic structure calculations and synthesis. Necessary requirement for this fundamentally physics based supervised machine learning approach is subdividing superconducting materials into classes and training on materials that share essential structural elements. We envisage this methodology to be extendable to all classes of superconducting materials. In parallel, disentangling the co-opetive interplay between features that become decisive for the mechanism for the superconductivity is work in progress.

**References**

1. F. London, and H. London, *Proceedings of the Royal Society A: Mathematical, Physical and Engineering Sciences*. **149**(866), 71, (1935).
2. J. Bardeen, L.N. Cooper, J.R. Schrieffer., *Phys. Rev.* **108**(5):1175–1204 (1957)
3. V.L. Ginzburg, L.D. Landau, *J. Exp. Theor. Phys.* **20**, 1064 (1950)
4. B. D. Josephson, *Phys. Lett.* **1**(7): 251–253 (1962).
5. A.A. Abrikosov, *Zh. Eksp. Teor. Fiz.* **32**, 1442 (1957)
6. Chu, C. W., Deng, L. Z. & Lv, B. Hole-doped cuprate high temperature superconductors. Physica C. **514**, 290–313 (2015).
7. Paglione, J. & Greene, R. L. High-temperature superconductivity in iron-based materials. Nat. Phys. **6**, 645–658 (2010).
8. Stewart, G.R., Physica C, , **514**, 28-35, (2015).
9. See e.g., P. P. Kong, V. S. Minkov, M. A. Kuzovnikov, Superconductivity up to 243 K in the yttrium-hydrogen system under high pressure, Nature Commun. 12, 5075 (2021)
10. W.E. Pickett, arXiv:2204.05930
11. Xie, S.R., Quan, Y., Hire, A.C. *et al.* Machine learning of superconducting critical temperature from Eliashberg theory. *npj Comput Mater* **8,** 14 (2022).
12. V. Stanev, C. Oses, A.G. Kusne, et al. *npj Comput Mater* **4**, 29 (2018).
13. P. G. de Gennes, *Superconductivity of Metals and Alloys* (Benjamin, New York, 1966).
14. L. N. Oliveira, E. K. U. Gross, and W. Kohn, *Phys. Rev. Lett.* **60,** 2430 (1988)
15. W. Kohn, W, EKU Gross, and LN Oliveira, *Int. J. of Quant. Chem.,* **36**(23), 611-615 (1989).
16. M. B. Suvasini, W. M. Temmerman, and B. L. Györffy, *Phys. Rev. B* **48**, 1202 (1993)
17. W. Kohn and J. M. Luttinger, *Phys. Rev. Lett*. **15**, 524, (1965).
18. I Panas*, J. Phys. Chem. B*, **103**(49), 10767–10774 (1999).
19. I Panas *Physica C: Superconductivity*, *480,* Pages 137-143 (2012)
20. I Panas, *Phys.Rev. B* 83 (2), 024508 (2011).
21. I Panas, A Snis, *Theoretical Chemistry Accounts* **97** (1), 232-239 (1997)
22. I Panas, *Molecular Physics* **89** (1), 239-246 (1996)
23. W. Kohn, and J.L. Sham, *Phys. Rev.* **140** (4A): A1133–A1138 (1965).
24. P. Hohenberg; W. Kohn *Phys. Rev.* **136** (3B): B864–B871 (1964).
25. D. Dew-Hughes, *Cryogenics* **15**(8), Pages 435-454 (1975)
26. K. M. Ho, W. E. Pickett, and M. L. Cohen, *Phys. Rev. B* **19**, 1751, (1979)
27. L. F. Mattheiss, *Phys. Rev. B* **12**, 2161 (1975).
28. C Paduani, *Brazilian Journal of Physics* **37**, 1073-107 (2007)
29. W. E. Pickett, K. M. Ho, and M. L. Cohen, *Phys. Rev. B* **19**, 1734 (1979)
30. SuperCon database, http://supercon.nims.go.jp/English.html

31. D. Dew-Hughes, *Cryogenics* **15**(8), Pages 435-454 (1975)
32. J. P. Perdew, K. Burke, and M. Ernzerhof, *Phys. Rev. Lett*. **77**, 3865 (1996)
33. H. J. Monkhorst and J. D. Pack, *Physical Review B***13**, 5188 (1976).
34. Stewart J. Clark, Matthew D. Segall, Chris J. Pickard, Phil J. Hasnip, Matt I. J. Probert, Keith Refson and Mike C. Payne. *Zeitschrift für Kristallographie*: **220** (5-6), 567-570 (2005).
35. MS. Materials Studio 6.0 *Accelrys Inc.*
36. F. Galasso, B. Bayles, and S. Sohele. *Nature* **198**, 984 (1963).
37. R.M. Waterstrat, J. Müller, *Journal of the Less Common Metals*, **52**(2), 271-277 (1977)
38. D. Dew-Hughes and V. D. Linse, *Journal of Applied Physics* **50**, 3500 (1979)
39. Jinhyuk Lim *et al* *J. Phys.: Condens. Matter* 33 285705 (2021)

Supplementary Information

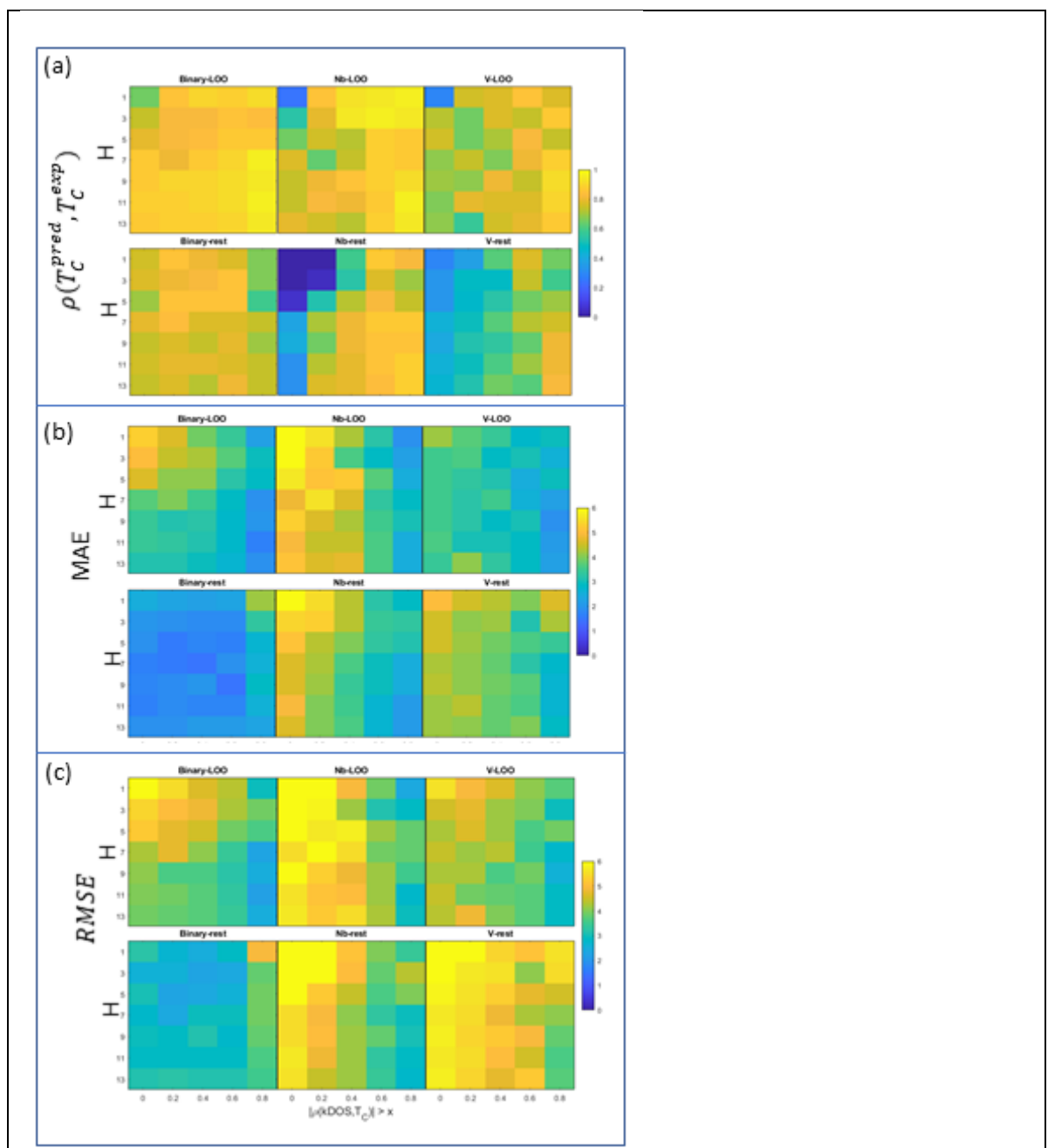

**Figure S1.** Heat maps of resulting (a) Pearson correlation coefficient $|\rho(T_C^{pred}, T_C^{exp})|$, (b) MAE, and (c) RMSE on the H-$|\rho(kDOS, T_C)|$ parameter space for regressions training on the all-binaries subset (left column), $Nb_3X$ (center column), and $V_3X$ (right column). Upper row corresponds to LOOCV within the specified subset and the lower row corresponds to regression predictions of a model trained on the specified subset on the members of the remaining subsets.

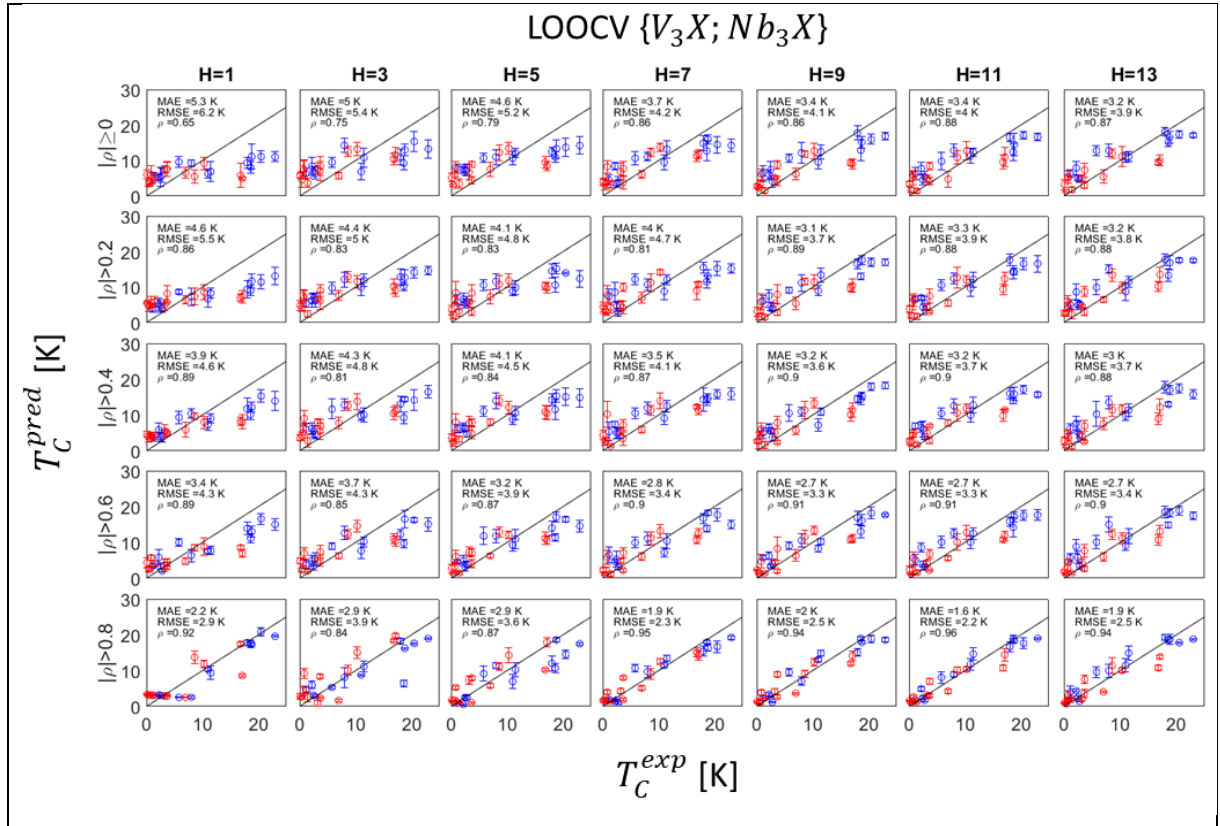

**Figure S2.** All-binaries subset: Leave-one-out regression analysis, showing the ability to predict $T_C$ of members in the subset based on training on all but one of the members. Rows, top down: Impact on prediction owing to filtering on kDOS features (pixels) that display correlation $|\rho(kDOS, T_C)| > 0$ to $0.8$. Columns, left to right: Impact of mean filtering using a square kernel of size H.

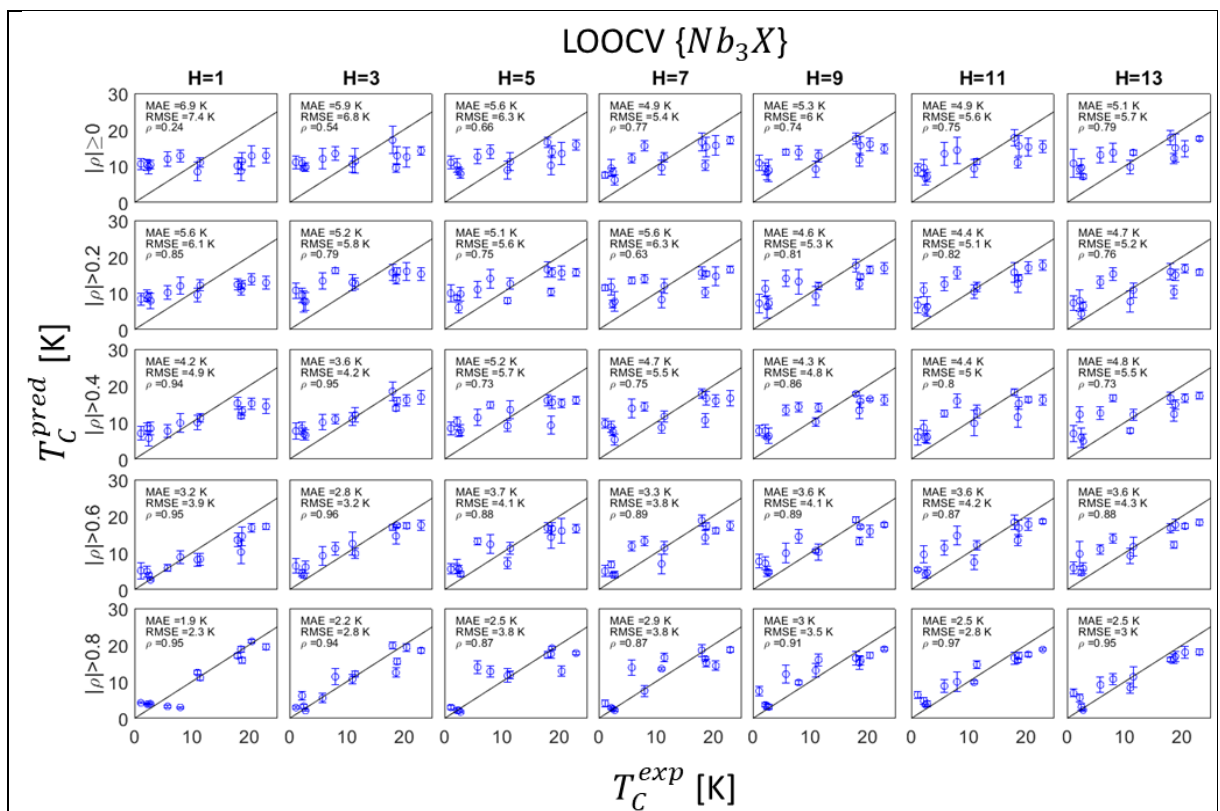

**Figure S3.** The Nb$_3$X subset: Leave-one-out cross-validation results, showing the ability to predict T$_C$ of members within the subset based on training on all but one of the members. Rows, top down: Impact on prediction owing to filtering on kDOS features (pixels) that display correlation $|\rho(kDOS, T_C)| > 0$ to 0.8. Columns, left to right: Impact of mean filtering using a square kernel of size H.

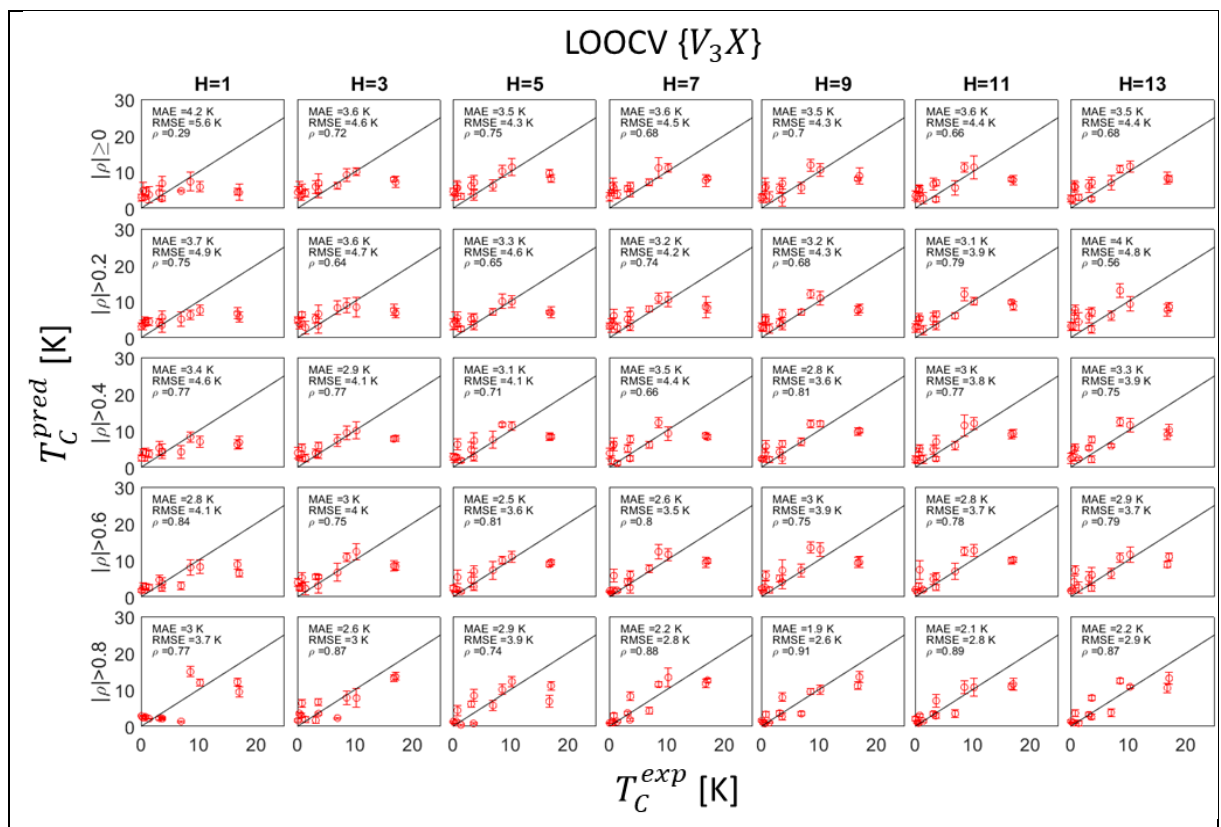

**Figure S4.** The $V_3X$ subset: Leave-one-out cross-validation results, showing the ability to predict $T_C$ of members in the subset based on training on all but one of the members. Rows, top down: Impact on prediction owing to filtering on kDOS features (pixels) that display correlation $|\rho(kDOS, T_C)| > 0$ to 0.8. Columns, left to right: Impact of mean filtering using a square kernel of size H.

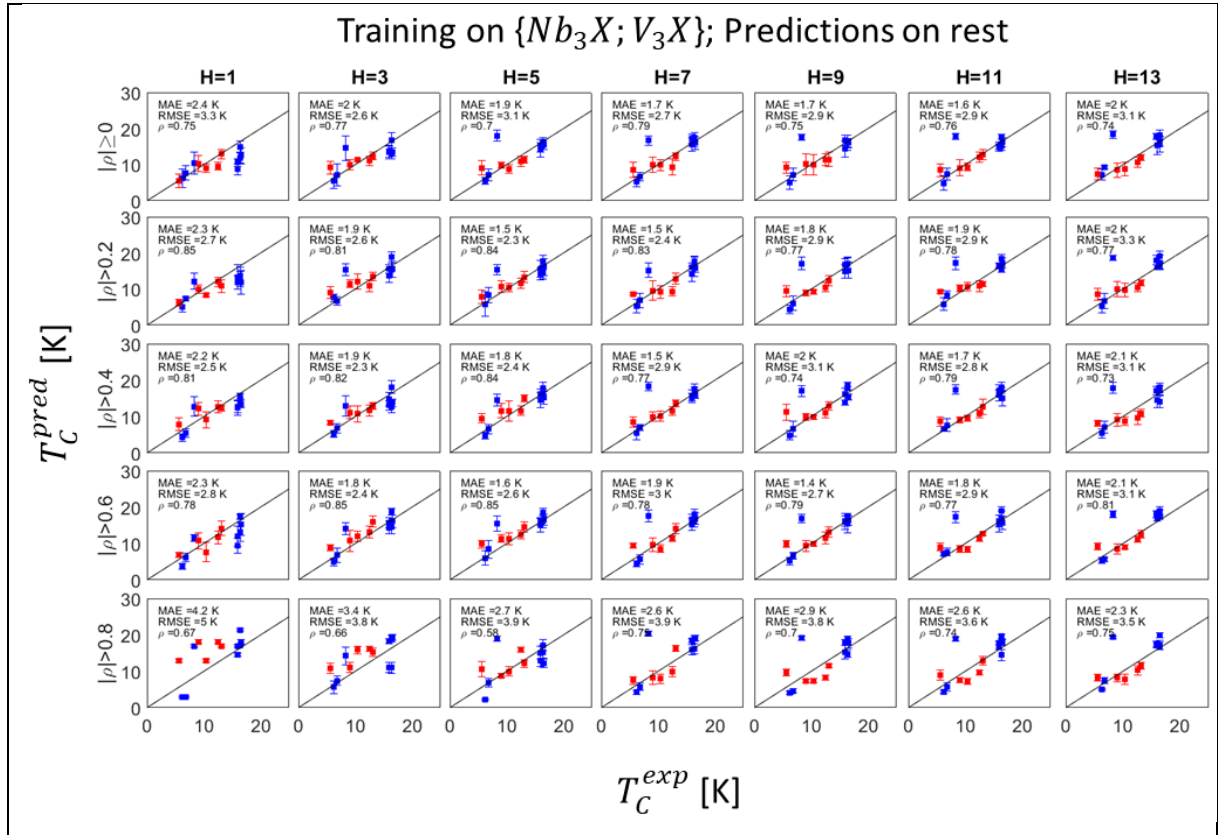

**Figure S5.** Predictions of a model trained on the all-binaries subset on $T_C$ in the ternary $A_6XY$, A=V (red), Nb (blue). Rows, top down: Impact on prediction owing to filtering on kDOS features (pixels) that display correlation $|\rho(kDOS, T_C)| > 0$ to 0.8. Columns, left to right: Impact of mean filtering using a square kernel of size H.

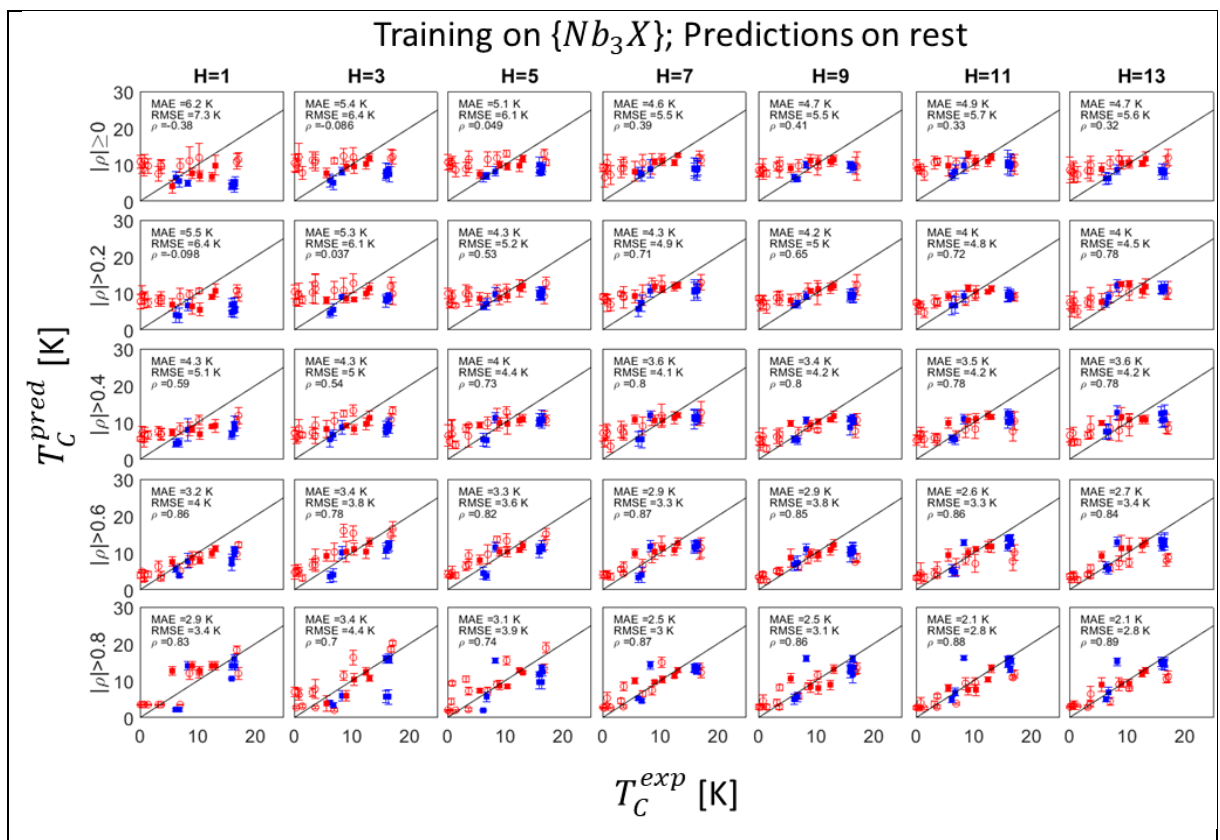

**Figure S6.** Predictions of a model trained on the Nb$_3$X subset on T$_C$ in the V$_3$X (red circles) and ternary A$_6$XY (filled squares) subsets, A=V (red), Nb (blue). Rows, top down: Impact on prediction owing to filtering on kDOS features (pixels) that display correlation $|\rho(kDOS, T_C)| > 0$ to 0.8. Columns, left to right: Impact of mean filtering using a square kernel of size H.

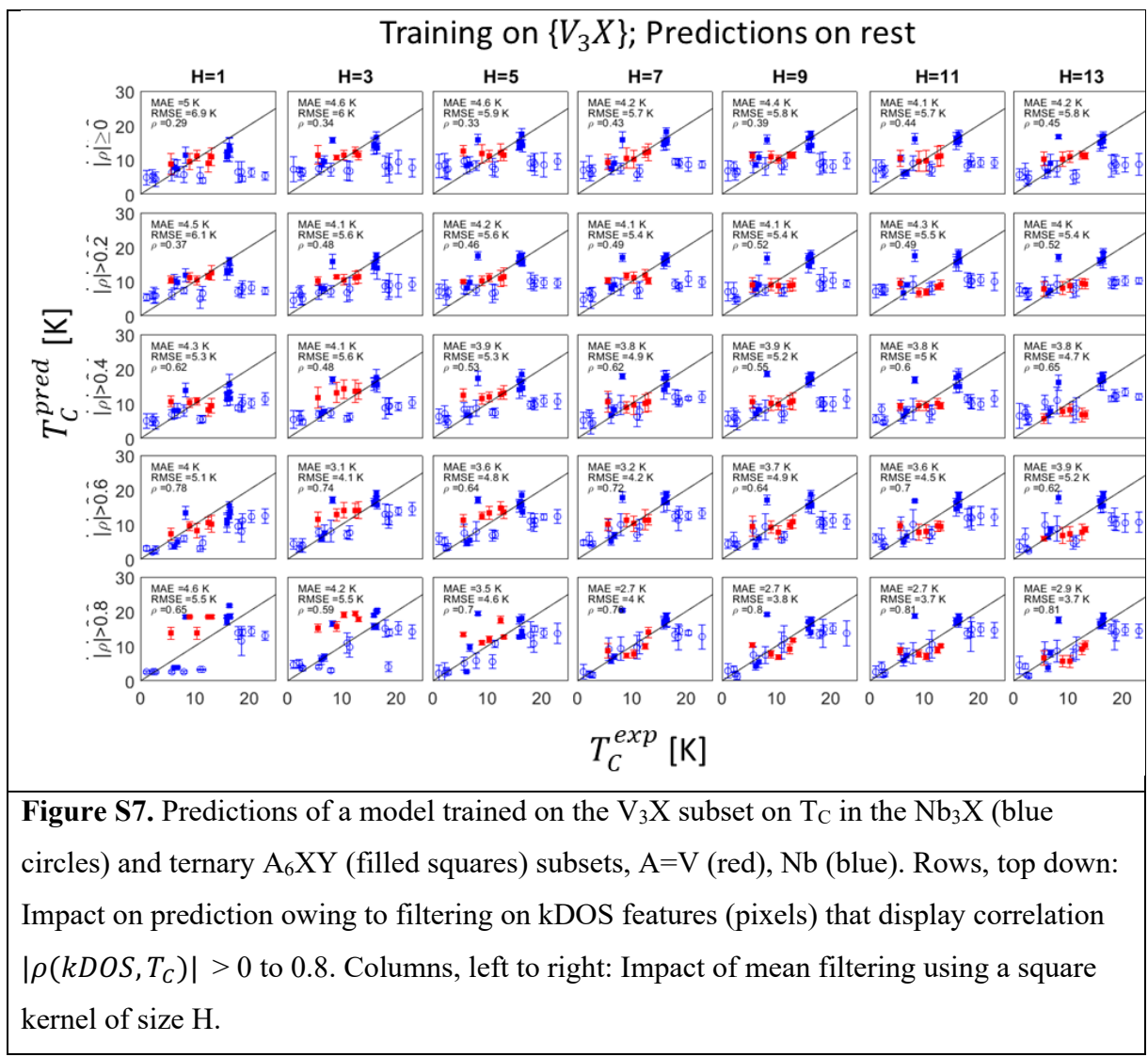

**Figure S7.** Predictions of a model trained on the V$_3$X subset on T$_C$ in the Nb$_3$X (blue circles) and ternary A$_6$XY (filled squares) subsets, A=V (red), Nb (blue). Rows, top down: Impact on prediction owing to filtering on kDOS features (pixels) that display correlation $|\rho(kDOS, T_C)| > 0$ to 0.8. Columns, left to right: Impact of mean filtering using a square kernel of size H.

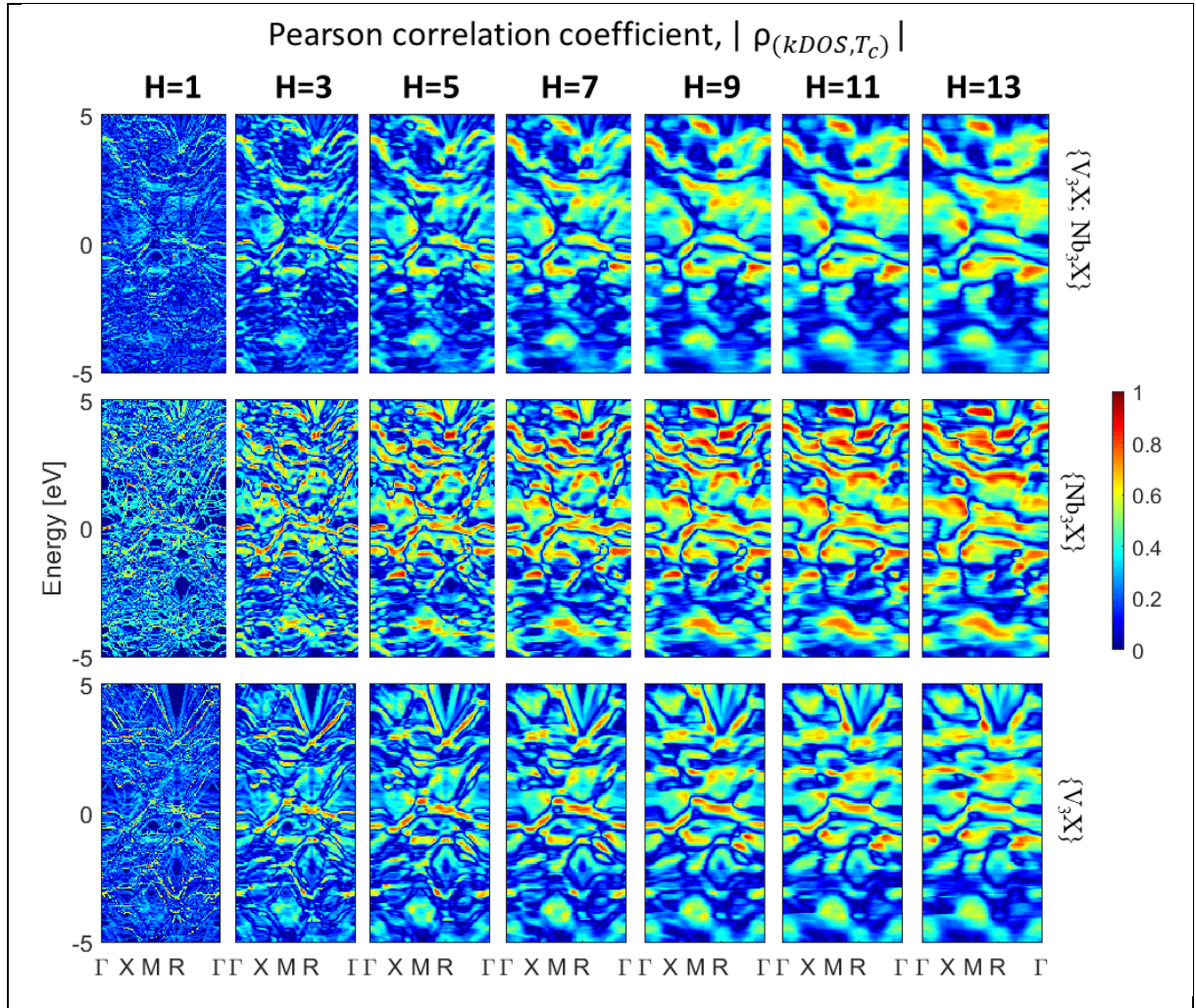

**Figure S8.** Sequence of the absolute value of Pearson correlation coefficient that result from all binaries (top), Nb$_3$X (middle), and V$_3$X (bottom). Columns, left to right, show the impact of mean filtering using a square kernel of size H on the correlation analysis.